\documentclass[12pt]{iopart}

\usepackage{iopams}  
\usepackage[final]{epsfig}
 
\begin{document}

\title[]{Systematics of parton-medium interaction from RHIC to LHC}
 
\author{\underline{Thorsten Renk}, Jussi Auvinen, Kari~J.~Eskola, 
Ulrich Heinz, Hannu Holopainen, Risto Paatelainen and Chun Shen }
 
\address{Department of Physics, P.O. Box 35, FI-40014 University of 
             Jyv\"askyl\"a, Finland; Helsinki Institute of Physics, P.O. Box 64, FI-00014 University 
             of Helsinki, Finland; Department of Physics, The Ohio State University, Columbus, 
             OH 43210, USA}
\ead{thorsten.i.renk@jyu.fi}
\begin{abstract}
Despite a wealth of experimental data for high-$P_T$ processes in heavy-ion 
collisions, discriminating between different models of hard parton-medium 
interactions has been difficult. A key reason is that the pQCD parton 
spectrum at RHIC is falling so steeply that distinguishing even a 
moderate shift in parton energy from complete parton absorption is 
essentially impossible. In essence, energy loss models are effectively 
only probed in the vicinity of zero energy loss and, as a result, at 
RHIC energies only the pathlength dependence of energy loss offers some 
discriminating power. At LHC however, this is no longer the case: Due to 
the much flatter shape of the parton $p_T$ spectra originating from 
2.76 $A$\,GeV collisions, the available data probe much deeper into the 
model dynamics. A simultaneous fit of the nuclear suppression at both RHIC 
and LHC energies thus has great potential for discriminating between 
various models that yield equally good descriptions of RHIC data alone.
\end{abstract}
 
 
\section{Introduction}
 
High-$P_T$ processes taking place in the background of the medium produced in ultrarelativistic heavy-ion (A-A) collisions are a cornerstone of the experimental A-A program at the LHC. The aim is to use these processes to do 
``jet tomography'', i.e. to study both the short-distance physics of 
the bulk medium (i.e. its relevant degrees of freedom) and 
the geometry of its expansion.
However, so far attempts to extract even solid qualitative statements about the nature of parton-medium interaction have not been successful. Two main reasons can be identified: 1) 
For steeply falling primary parton spectra, medium-induced shifts in parton energy cannot be distinguished from parton absorption, hence observables lose sensitivity to model 
details; 2) Computed observable quantities depend both on assumptions made about the bulk medium evolution and the parton-medium interaction. In this work, we propose to resolve the second ambiguity by a systematic investigation of multiple observables, while we demonstrate that the first is significantly lessened for the harder LHC parton kinematics.
 
\section{Dependence on medium modelling}
 
We test different combinations of medium evolution and parton-medium interaction models against a large body of high $P_T$ observables. In particular, for the medium evolution we use a 3+1d ideal \cite{hyd3d}, a 2+1d ideal \cite{hyd2d,hydEbyE}
and a 2+1d viscous hydro code \cite{vhyd} with both CGC and Glauber initial conditions. On the parton-medium interaction side, we test a radiative energy loss model \cite{ASW}, a parametrized \cite{Elastic} and a Monte-Carlo (MC) model \cite{ElasticMC} for incoherent energy loss, a strong-coupling phenomenological model based on AdS/CFT ideas \cite{AdS} and the MC in-medium shower code YaJEM \cite{YaJEM1,YaJEM2} with its variant YaJEM-D \cite{YaJEM-D} which introduces an explicit pathlength/energy dependence into the minimum virtuality scale down to which the shower is evolved.
 
\begin{figure}[htb]
\epsfig{file=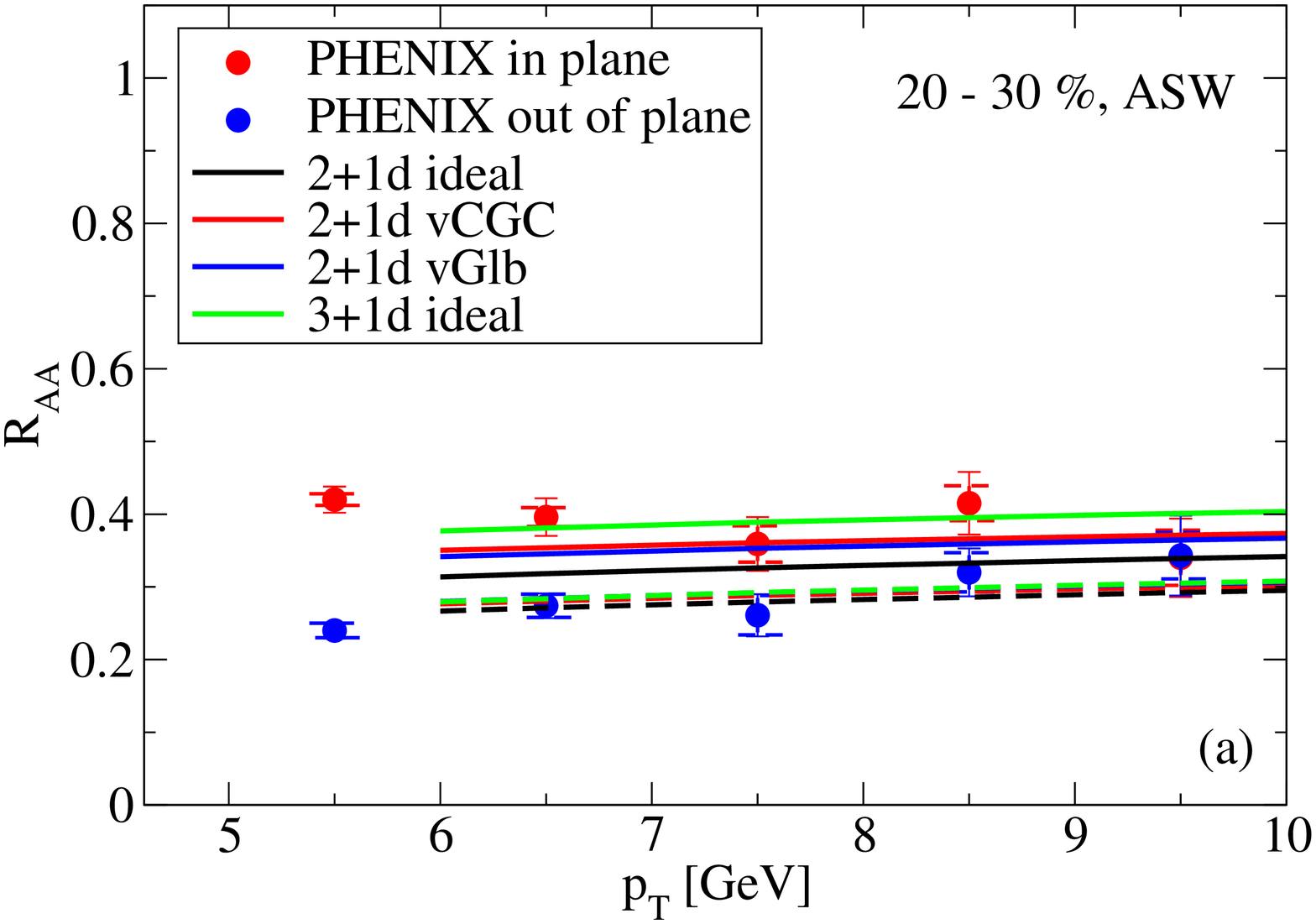, width=7.8cm}\epsfig{file=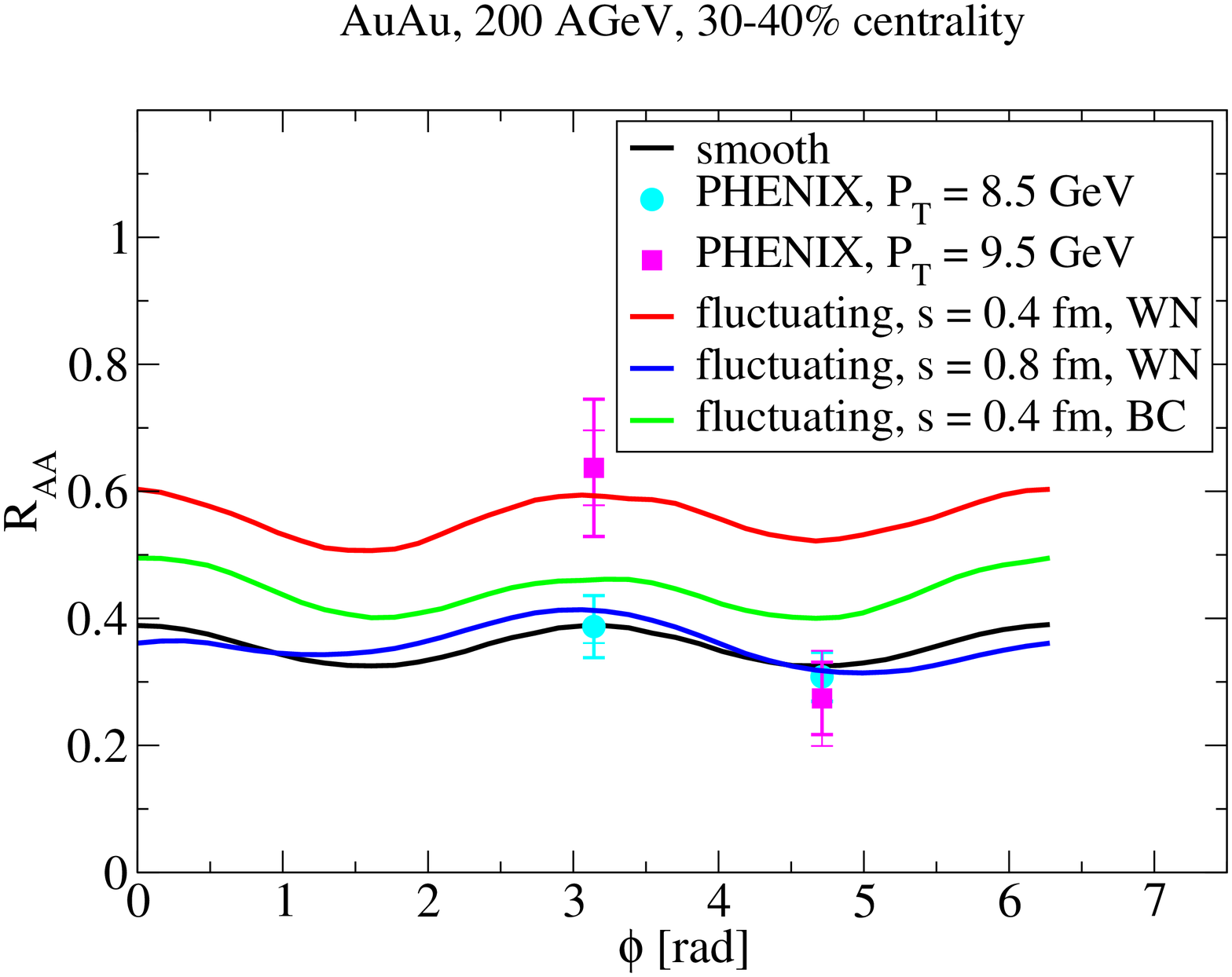, width=7.8cm}
\caption{\label{F-1}Left panel: $R_{AA}(P_T)$ for 30-40\% cental 200 AGeV Au-Au collisions for in plane (solid) and out of plane (dashed) emission computed for the same energy loss model (ASW) with different hydrodynamical backgrounds, compared with PHENIX data \cite{PHENIX_RAA_phi}. Right: $R_{AA}(\phi)$ at $P_T=10$ GeV for smooth, initial-state averaged and event-by-event fluctuating hydrodynamics for different fluctuation size scale.}
\end{figure}
 
In a first run, we pose the question to what degree the underlying medium model is able to influence high $P_T$ observables. In Fig.~\ref{F-1} we present an example of results from a systematic investigation of both the influence of smooth hydrodynamical models \cite{JetHydSys} and event-by-event hydrodynamics \cite{hydEbyE} with initial state density fluctuations \cite{JetFluct}. Summarizing the result, we find that the medium model has a considerable (factor $\sim 2$) influence on observables such as $v_2$ at high $P_T$ or extracted parameters such as the transport coefficient $\hat{q}$. Chiefly responsible is the location of the freeze-out hypersurface --- the agreement with data in general improves if the freeze-out hypersurface is large, but noticeable effects are also caused by the initialization time or the presence/absence of viscosity. Fluctuations in the hydrodynamical initial state play a minor role for extracted parameters ($\sim$ 20\%), due to a cancellation of competing effects (see \cite{JetFluct}), but for non-central collisions the cancellation is incomplete, leading to a decrease in suppression for small fluctuation size scale. If hard probes are used to constrain the fluctuation size, a scale of $\sim 0.8$ fm is preferred.
 
\section{Pathlength dependence}
 
In a second run, we test the pathlength dependence of the available parton-medium interaction models against the data \cite{YaJEM-D,JetHydSys}. In a static medium with constant density, we expect incoherent processes to scale with pathlength $L$ (elastic), radiative energy loss (ASW) with $L^2$ due to coherence effects and the strong coupling model (AdS) with $L^3$ due to the drag-force like interaction of the virtual gluon cloud with the medium. The shower code YaJEM is known to have in principle an $L^2$ dependence due to coherence, but which effectively reduces to $L$ by finite-energy corrections, whereas YaJEM-D has a complicated non-linear pathlength dependence. In an evolving hydrodynamical medium, the pathlength dependence is effectively much more complicated due to effects like spatial inhomogeneities in the medium, longitudinal and transverse flow and viscous entropy production.
We find that the data allow to unambiguously rule out linear pathlength dependence, leading to the conclusions that a large component ($>10$\%) of elastic energy loss is not favoured by the data and that finite-energy 
corrections to coherence arguments need to be taken seriously. The other models we tested (ASW, AdS and YaJEM-D) remain viable with the data, 
although in each case only in combination with a particular hydrodynamical evolution model.
 
\section{Extrapolation to large $\sqrt{s}$}
 
Using the EKRT saturation model, we can extrapolate one of our default hydro runs to LHC energies \cite{RAA_LHC} and thus 
significantly reduce the uncertainty in the hydrodynamical modelling, as the well-defined extrapolation procedure allows to compare results for 'the same' hydrodynamics at different $\sqrt{s}$.
 
\begin{figure}[htb]
\vspace*{-0.3cm}
\epsfig{file=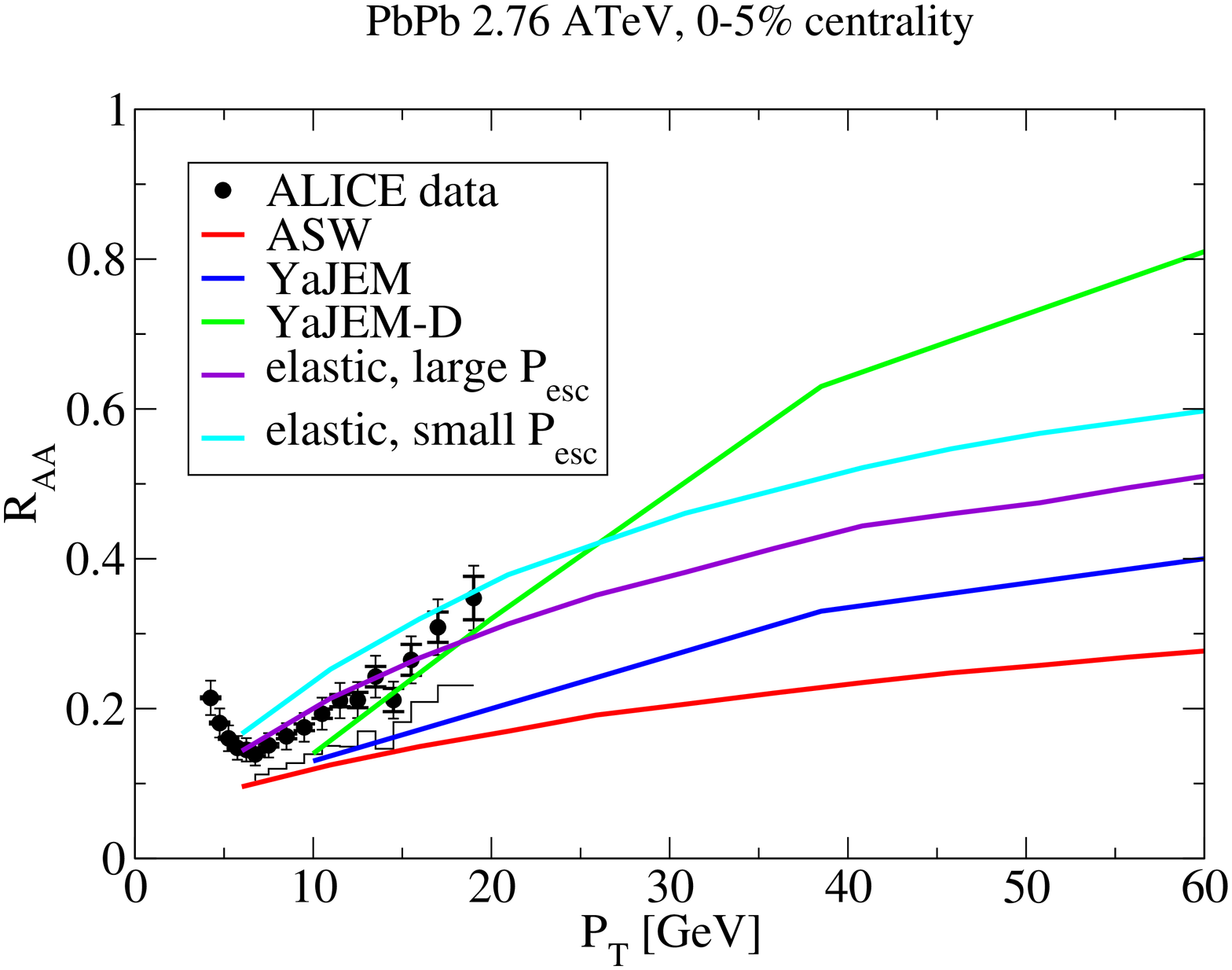, width=7.7cm}\epsfig{file=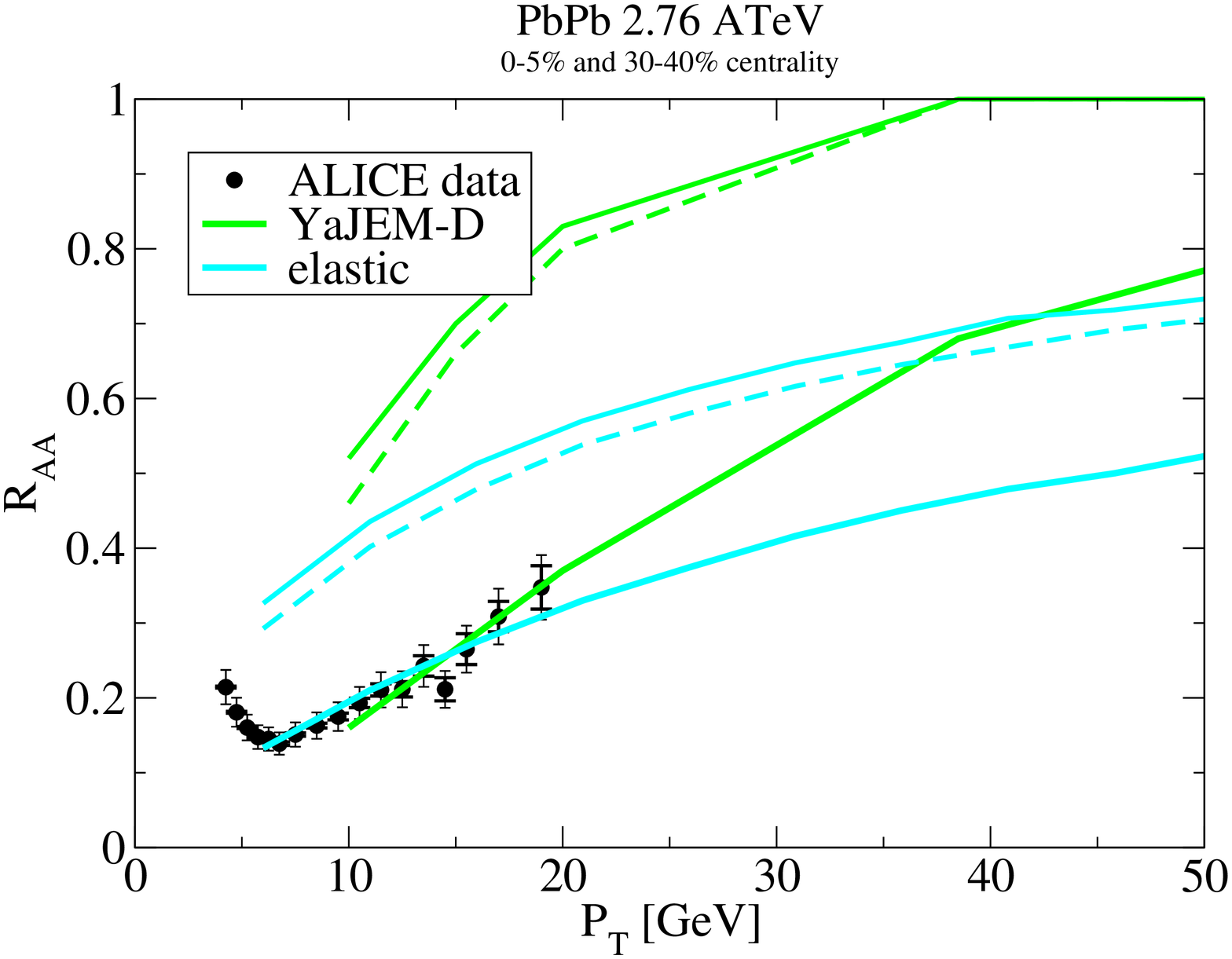, width=7.7cm}
\caption{\label{F-LHC}Left: The nuclear suppression factor $R_{AA}$ extrapolated from best fits at RHIC to central Pb-Pb collisions at LHC for various models (see text) compared with ALICE data \cite{ALICE}. Right: $R_{AA}$ for non-central collisions in-plane and out of plane based on a best fit to central LHC data.}
\end{figure}
 
Fig.~\ref{F-LHC} shows results for $R_{AA}$ in central and non-central collisions.
As expected, the sensitivity to model details in the $P_T$ dependence of the 
nuclear suppression factor is found to be much larger at LHC than at RHIC.
Combined with pathlength-dependent observables such as $R_{AA}(\phi)$ or the dihadron suppression factor $I_{AA}$, precision LHC data thus 
have a high potential for distinguishing between different models.
 
We can quantify the quality of extrapolation in $\sqrt{s}$ by the factor $R$ which quantifies the differences of parton-medium interaction parameters for a best fit to RHIC and LHC 
data, using the same hydrodynamical models. $R=1$ indicates an extrapolation without any tuning. 
We find that the shower codes 
YaJEM-D ($R=0.92$) and YaJEM ($R=0.61$) 
extrapolate reasonably well, whereas the radiative energy loss scenario 
ASW ($R{\,=\,}0.47$) is not favoured by the data, and a strongly coupled scenario AdS is strongly disfavoured with $R=0.31$. The latter effect can be readily understood by a dimensional analysis --- if the pathlength dependence is $L^3$, then for dimensional reasons the model must probe the medium temperature as $T^4$, i.e. the model
responds much more strongly than all others to the higher initial medium density at the LHC, which leads to overquenching. Given this finding, there is currently no reason to assume that the data would prefer a strongly coupled over a perturbative scenario of parton-medium interaction.
 
Combining the constraints from pathlength dependence and $\sqrt{s}$ 
extrapolation, assuming no systematic uncertainty on the published LHC 
data \cite{ALICE}, out of the models tested here only YaJEM-D together 
with a hydrodynamics similar to the 3+1d code remains a viable description. 
We take this as a strong indication that systematic studies along the 
lines discussed here are indeed a suitable tool to do jet tomography.
 
 
\vspace*{0.3cm}
 
{\bf Acknowledgments:} 
This work was supported by the Finnish Academy (projects 130472 and 133005), 
the Finnish Graduate School of Particle and Nuclear Physics, the Jenny and Antti Wihuri Foundation, the Magnus 
Ehrnrooth Foundation and the U.S. Department of Energy (grants
\rm{DE-SC004286} and (within the framework of the JET Collaboration) 
\rm{DE-SC0004104}).

\end{document}